\documentclass{iopjournal}
\usepackage{lmodern}
\usepackage{float}
\usepackage{amsmath,amssymb,graphicx}  
\usepackage{upgreek}
\usepackage{hyperref}
\usepackage{cite}
\usepackage{physics}
\usepackage{xcolor}
\usepackage[font=footnotesize]{caption}
\usepackage{subcaption}
\usepackage{circuitikz}
\usepackage{relsize}
\usepackage{nicematrix}
\newcommand{\dms}[1]{\hat{\uprho}_{Q}(#1)}

\begin{document}

\articletype{Paper} 

\title{Non-Markovian Charging of a Micromaser Quantum Battery}

\author{Nicolò Massa$^{1}$\orcid{0009-0005-7092-6209}, Dario Ferraro$^{1,2,*}$\orcid{0000-0002-4435-1326}, Fabio Cavaliere$^{1,2}$\orcid{0000-0000-0000-0000}, Giuliano Benenti$^{3,4}$\orcid{0000-0000-0000-0000} and Maura Sassetti$^{1,2}$\orcid{0000-0000-0000-0000}}

\affil{$^1$Dipartimento di Fisica, Universit\`a di Genova, Via Dodecaneso 33, I-16146 Genova,~Italy}

\affil{$^2$CNR-SPIN, Via Dodecaneso 33, I-16146 Genova,~Italy}

\affil{$^3$Center for Nonlinear and Complex Systems, Dipartimento
di Scienza e Alta Tecnologia, Universit\`a degli Studi dell'Insubria,
via Valleggio 11, 22100 Como, Italy}

\affil{$^4$Istituto Nazionale di Fisica Nucleare, Sezione di Milano,
via Celoria 16, 20133 Milano, Italy}

\affil{$^*$Author to whom any correspondence should be addressed.}

\email{dario.ferraro@unige.it}


\begin{abstract}
We theoretically investigate the role played by non-Markovianity in the charging of a quantum battery. We focus on the case of a Micromaser, where the energy is stored in a resonant cavity mode and memory effects are introduced and controlled by coupling this mode to a memory ancilla, described as a two-level system, that in turn sequentially interacts with a stream of two-level chargers through pairwise exchange interactions leading to a partial swap. Through numerical analysis, we show that non-Markovian memory effects act as a resource that enhances both the charging capacity and the stability of this device with respect to the Markovian counterpart. We conclude our study showing that the proposed scheme can be implemented in state-of-the-art solid-state platforms for circuit quantum electrodynamics.  
\end{abstract}


\section{Introduction}
The progressive development of technologies based on quantum mechanics\cite{Ezratty24, Aguado24} triggered an intense research activity addressing the design and 
realization of devices capable of storing, transferring and manipulating energy at the micro and nano scale. These systems, known as quantum 
batteries (QBs), can be promoted from the ground state to highly excited states exploiting genuinely quantum mechanical features such as coherence, entanglement and collective 
interactions~\cite{Quach23, Campaioli24, Ferraro26}. Since the pioneering theoretical proposals of QBs based on collections of two-level systems~\cite{Alicki13, Binder15, Campaioli17}, a growing body of fundamental and applied  
research has investigated charging power~\cite{Gyhm22, Safranek23}, energy storage capacity~\cite{Farre20, Cavaliere26} and fluctuations~\cite{Friis18, Rinaldi25, Donelli25} in a variety of
physical platforms including spin systems~\cite{Le18, Joshi22, Cruz22, Grazi24, Catalano24, Lu25, Grazi25, Cenedese26, Varrica2026}, cavity quantum electrodynamics setups~\cite{Ferraro18, Ferraro19, Quach22, Gemme23, Erdman24, Beder25, Hymas25, Canzio25}, superconducting circuits~\cite{Hu22, Gemme24, Cavaliere25, Razzoli25} and trapped ions~\cite{Zhang25}. 

In this context, collisional models (CMs) have emerged as a useful framework for describing the charging and 
discharging processes of QBs~\cite{Seah21, Shaghaghi_MM_22, Shaghaghi_MM_23, Salvia23, Rodriguez23, Massa25, Crotti26}. These models were originally introduced in the context of open quantum systems to provide an effective description of the coupling to an external dissipative environment, by decomposing the latter into elementary constituents~\cite{Ciccarello21, Morrone23, Elyasi2025,Li26}. The key idea behind CMs is to assume that the complex open dynamics can be described as a sequence of simpler interactions involving the system and a single environmental building block at a time.
Moreover, they are closely connected with 
experimentally relevant scenarios involving repeated quantum interactions, such as the Nobel prize awarded Micromaser
experiments carried out by Serge Haroche and coworkers~\cite{Deleglise2008,Haroche13}. 

Differently from what happens for the conventional dissipative case~\cite{Ciccarello21, Morrone23, Wang2026}, in collisional QBs setups the environment becomes active, with ancillas playing the role of chargers, whose interactions with the cavity are suitably engineered to efficiently transfer energy into it. 
In particular, in a Markovian Micromaser QB, a resonant cavity mode described as a quantum harmonic oscillator interacts sequentially with a stream of identical and independent ancillary two-level quantum systems, forming what is known in literature as a $\textit{quantum current}$ ~\cite{Slosser_89,Seah21, Shaghaghi_MM_22, Shaghaghi_MM_23, Salvia23, Crotti26}. Under proper conditions, each individual collision realize a net energy transfer from the charger to the QB, allowing 
its progressive charging. Within this framework, Refs.~\cite{Shaghaghi_MM_22, Shaghaghi_MM_23} showed that a flow of ancillas prepared in identical quantum states with non-zero coherences at the level of the density matrix can lead to a smaller, but more stable and controllable, stored energy with respect to the case where such coherences are not present.

The assumption of a Markovian dynamics greatly simplifies the analytical and numerical description of the problem, but neglects memory effects that are ubiquitous in realistic conditions~\cite{Rivas14, Breuer16,Li25}. In general, these effects are expected to induce deviations from the Markovian behavior and can have a substantial impact also at the level of the charging dynamics of QBs~\cite{Cavaliere25}. Therefore, understanding the interplay between memory effects and 
energy transfer is of fundamental importance to characterize more realistic energy storage devices. In this direction, CMs provide
an ideal setting for exploring the role played by non-Markovianity in charging protocols. Within this framework, a possible way to introduce memory effects consists in allowing the environmental constituents to interact with one another, breaking the independent ancillas assumption and thereby retaining information from previous collisions~\cite{Ciccarello21}.  

While, as stated above, Micromaser QBs have been extensively studied within Markovian collisional frameworks, the impact of controlled memory effects on their charging performance remains largely unexplored.
In the present paper, we will fill this gap by investigating the role played by non-Markovian effects in collisional QBs setting. To this end, we consider a memory-assisted Micromaser QB, where a cavity mode is coupled to a dedicated memory ancilla mediating the energy transfer from a stream of chargers. Such kind of scheme can be easily handled numerically and is equivalent to the case where a flow of interacting chargers is coupled directly to the harmonic oscillator. We will address both the step by step dynamics and the steady state of the system, showing that memory effects not only increase the energy stored into the battery but also substantially suppress energy fluctuations, thus improving the reliability of the energy storage process~\cite{Mohan26}. This occurs at the expense of a slowing down of the charging. The possibility to realize the considered scheme in state of the art superconducting circuits~\cite{Krantz19} will also be discussed to further strengthen our proposal.

The paper is organized as follows. Section \ref{sec:model} discusses the building blocks composing the memory assisted Micromaser QB under investigation and the proposed non-Markovian charging scheme. The relevant figures of merit considered to characterize the functioning and the reliability of the QB are also introduced. In Section \ref{sec:results} we discuss the results concerning the effects of non-Markovianity on the charging of the Micromaser QB. The experimental feasibility of the proposed device is discussed in Section \ref{sec:exp_feas}.


\section{Model}
\label{sec:model}


\subsection{Memory assisted Micromaser QB}
\label{subsec:Mem_assisted_MM}

The QB setup under investigation consists of a single cavity radiation mode playing the role of the battery itself. It is charged via interaction with an auxiliary quantum system, the memory ancilla. The latter in turn interacts sequentially with other ancillas playing the role of chargers.\\
From now on, the radiation mode will be denoted by $Q$ and will be described as a quantum harmonic oscillator of frequency $\omega$ with Hamiltonian ($\hbar=1$)
\begin{equation}
    \hat{H}_Q = \omega \hat{a}^\dagger \hat{a}
\end{equation}
with 
$\hat{a}$ ($\hat{a}^{\dagger}$) indicating its annihilation (creation) operator. Notice that here we have set the energy reference in such a way to neglect the (non-extractable) zero point energy of the oscillator.\\
The set of ancillas will be denoted by $\mathcal{A}=\qty{M, A_1, \dots A_N}$, where $M$ is the memory ancilla and $A_1\dots A_N$ are the $N$ external chargers. The elements of $\mathcal{A}$ will be described as identical two-level systems (TLSs) described by the Hamiltonians
\begin{equation}
    \hat{H}_k = \frac{\omega}{2}\hat{\sigma}^{(k)}_{z}
\end{equation}
where $\hat{\sigma}^{(k)}_{\alpha}$ ($\alpha=x,y,z$) are the conventional Pauli matrices for the $k$--th ancilla ($k\in \mathcal{A}$). We assume henceforth the following hypotheses: the energetically favorable condition in which both the QB and the TLSs have the same level spacing $\omega$ (resonant regime) and the homogeneity assumption of  identical ancillas.\\

\begin{figure}[h]
\centering
     \begin{subfigure}[c]{0.41\textwidth}
         \centering
         \includegraphics[width=\textwidth]{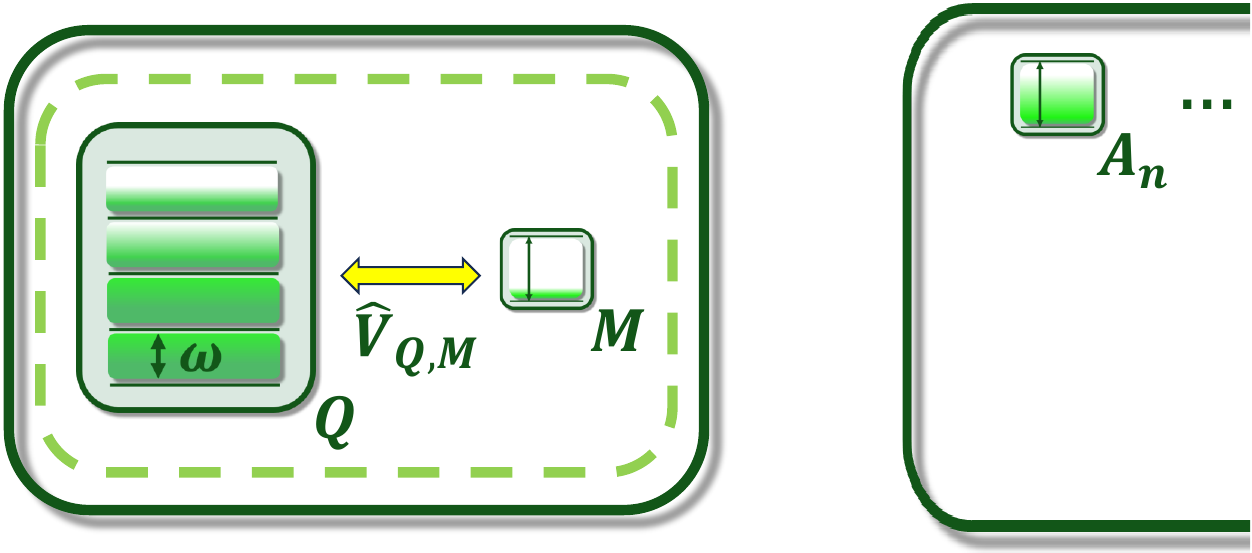}
         \caption{Step \emph{i})}
     \end{subfigure}
     \hspace{2cm}
     \begin{subfigure}[c]{0.41\textwidth}
         \centering
         \includegraphics[width=\textwidth]{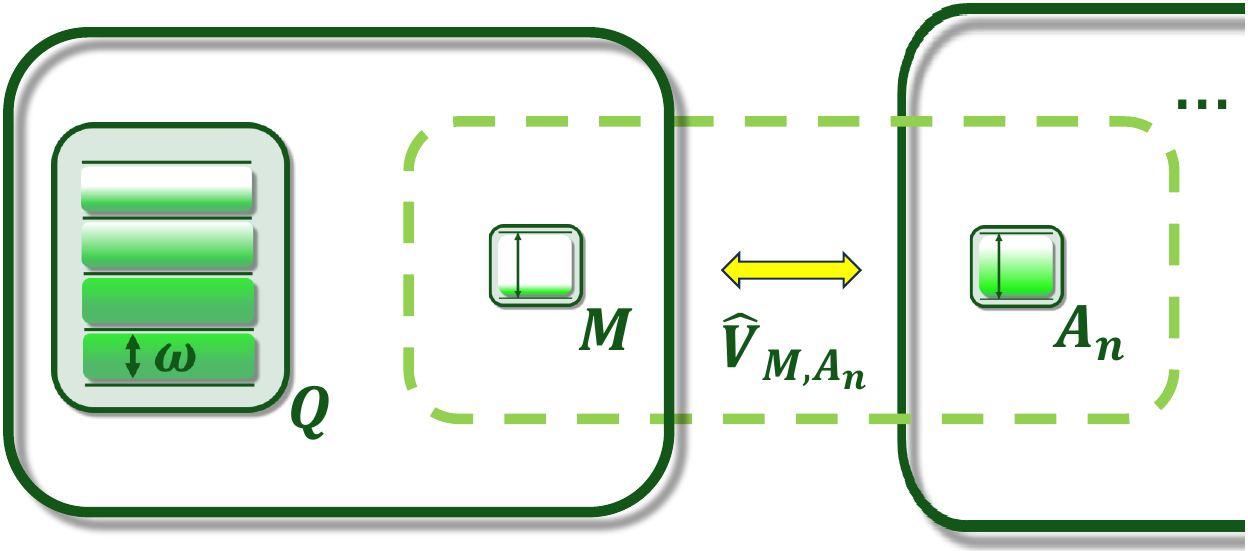}
         \caption{Step \emph{ii})}
     \end{subfigure}
     
     \caption{Scheme of the collisional step used for the charging protocol: (a) the battery $Q$, given by a harmonic oscillator with level spacing $\omega$, interacts with the memory ancilla $M$ via the coupling $\hat{V}_{Q, M}$; (b) $M$ interacts with the $n$-th charger $A_n$, via the term $\hat{V}_{M,A_{n}}$.}
     \label{fig:fig1}
\end{figure}

The charging scheme proposed in this work develops as follows: it starts with a $Q-M$ interaction, followed by the interaction of the memory with the first external charger $M-A_1$. After the interaction $A_1$ is sidelined; this twofold step is then repeated, with $Q-M$ followed by $M-A_2$, and so on. We will refer to each of these sequences $Q-M$, $M-A_n$ as a \textit{collisional step}: the overall process realizes a \textit{composite CM}~\cite{Ciccarello21}. Within this scheme, the memory ancilla emerges as the key component to introduce memory effects in our model, since it accumulates information coming from all the chargers it interacts with~\cite{Ciccarello21}. A cartoon view of this single \textit{collisional step} is reproduced in Fig.~\ref{fig:fig1}.\\

In what follows, the initial state of the QB, $\dms{0}$ (before the first collision), will be assumed as its ground state while the memory and chargers $\qty{\hat{\eta}_M,\hat{\eta}_{A_1},\dots,\hat{\eta}_{A_N}}$ ancillas will be initialized in the same superposition of ground and excited state, described by the density matrix
\begin{equation}
\label{eq:anc_dm}
    \hat{\eta}_{k}(0) = q\ketbra{0}{0}_k+(1-q)\ketbra{1}{1}_k+c\sqrt{q(1-q)}
    (\ketbra{1}{0}_k+\ketbra{0}{1}_k).
\end{equation} 
In the above expression the parameter $c$ allows us to encode both fully coherent ($c=1$) and completely incoherent ancillas ($c=0$), whereas $q$ is the population of the ancillas ground state~\cite{Shaghaghi_MM_22}.
The target $q$ regime in order to realize a significant charging of the battery is $q<0.5$, namely population inverted ancillas, as will be confirmed by our results.

\subsubsection{Battery-memory interactions}
The interaction between the QB and the memory $M$, (corresponding to step \emph{i} in Fig. \ref{fig:fig1}), is ruled by the dipolar matter-radiation interaction~\cite{Jaynes63, Larson24}
\begin{equation}
    \label{eq:interact_pot}
    \hat{V}_{Q,M} = g (\hat{a}^\dagger + \hat{a})\otimes\hat{\sigma}^{(M)}_{x}.
\end{equation} 
Notice that, in the above expression, the notation of the Pauli matrix keeps explicitly track of the fact that it involves the memory ancilla. 
Moreover, $g$ denotes the coupling strength and is assumed to be step independent, in agreement with the homogeneity requirement.\\
The dynamics is usually investigated, for sake of simplicity, in the interaction picture~\cite{Ciccarello21, Crotti26}. According to this, each $Q-M$ interaction is ruled by the unitary operator
\begin{equation}
\label{eq:unitary}
    \hat{U}_{Q,M}(\tau) = \mathcal{T}\bigl\{e^{-i\int_{0}^{\tau}\hat{V}_{Q,M}(t')\,dt'}\bigr\},
\end{equation}
with $\mathcal{T}$ indicating the time-ordering and where $\tau$ is the duration of each interaction. Also in this case the homogeneity assumption implies that the time steps are $n$-independent. Within the interaction picture, the Hamiltonian in Eq. (\ref{eq:interact_pot}) becomes
\begin{equation}
\label{eq:V_s_anc_full}
    \hat{V}_{Q,M}(t) = g\biggl[
    \hat{a}^\dagger \hat{\sigma}^{(M)}_{-} + \hat{a} \hat{\sigma}^{(M)}_{+} +
    \hat{a} \hat{\sigma}^{(M)}_{-}e^{-2i\omega t}+\hat{a}^\dagger \hat{\sigma}^{(M)}_{+} e^{2i\omega t}
    \biggr],
\end{equation}
where
\begin{equation}
\hat{\sigma}^{(M)}_{\pm}=\frac{\hat{\sigma}^{(M)}_{x}\pm i\hat{\sigma}^{(M)}_{y}}{2}.
\end{equation}
The first two contributions in Eq. (\ref{eq:V_s_anc_full}) are usually referred to as $\textit{rotating terms}$, whereas the time dependent ones as $\textit{counter-rotating (CR)}$.\\
The time evolution gets considerably simplified by neglecting the CR terms, an assumption known as $\textit{Jaynes-Cummings}$ (JC) limit or rotating wave approximation (RWA), whose accuracy strongly depends on the relative intensity of the coupling strength, namely  $g/ \omega$~\cite{Schleich, Krantz19}. Indeed, recent works on Micromaser QBs~\cite{Shaghaghi_MM_22, Huang_20} pointed out that CR terms start affecting noticeably the system's dynamics for $g/\omega \gtrsim 0.1$. Taking into account this constraint, in the following we will work in the $g/\omega < 0.1$ limit, neglecting the CR terms. Under this condition, Eq. (\ref{eq:unitary}) takes the simpler form
\begin{equation}
\label{eq:SM_unitary}
    \hat{U}_{Q,M}(\theta) = e^{-i\theta\bigl[\hat{a}^\dagger \hat{\sigma}^{(M)}_{-} + \hat{a}\hat{\sigma}^{(M)}_{+}\bigr]}
\end{equation}
where we have introduced the short notation
\begin{equation}
    \theta \equiv g\tau
\end{equation}
since no time-ordering is needed.\\

According to the above discussion, it is evident that the $Q-M$ dynamics only depends on the parameter $\theta$, which can determine two different behaviors (see Appendix \ref{Trapping} for more details). The first one, termed \textit{fine-tuned dynamics}, occurs if 
\begin{equation}
\label{eq:theta_ft}
    \theta =\theta_{ft} = \frac{\pi}{\sqrt{s}}, s \in \mathbb{N}^{*}
\end{equation}
and leads to the decomposition of the QB Hilbert space into a set of dynamically disconnected blocks (also indicated as \textit{traps}), causing the density matrix of the QB to be block-diagonal in the energy eigenbasis throughout the entire evolution~\cite{Slosser_89,Nemeth_05,Shaghaghi_MM_22,Shaghaghi_MM_23}. This implies that a QB initialized in the ground state (empty QB) will remain confined (trapped) within a block of dimension $s\times s$ in the Fock state space. \\
The situation is different for $\theta\neq \theta_{ft}$, which is referred to as
\textit{non fine-tuned dynamics}~\cite{Shaghaghi_MM_23}. Here, the block diagonal structure is broken and the battery is no longer trapped within the $s\times s$ block containing the ground state, and is therefore allowed to explore the region outside the trap.

\subsubsection{Memory-charger interaction}
\label{subsubsec:M_A_interactions}
Concerning the coupling between the memory and each charger in $\qty{A_1,A_2,\dots A_N}$, it is based on a pairwise exchange interaction with intensity $g_{a}$
\begin{equation}
\label{eq:aa_int_pot}
\hat{V}_{M,A_{n}} = \frac{g_{a}}{2}\hat{\vec{\sigma}}^{(M)}\cdot \hat{\vec{\sigma}}^{(A_{n})},
\end{equation}
with $\hat{\vec{\sigma}}^{(A_k)}=\left(\hat{\sigma}^{(A_k)}_{x},\hat{\sigma}^{(A_k)}_{y}, \hat{\sigma}^{(A_k)}_{z}\right)$ and where again the notation is chosen to make explicit the involved ancillas.
Moving to interaction picture, the coupling in Eq. (\ref{eq:aa_int_pot}) is formally left unchanged.  Denoting with $\tau_{a}$ the duration of such interaction, (corresponding to step \emph{ii} in Fig. \ref{fig:fig1}), one obtains the unitary evolution 
\begin{equation}
    \hat{W}_{M,A_n} = e^{-i \tau_{a} \hat{V}_{M,A_n}}=e^{-i \frac{\theta_{a}}{2} \hat{\vec{\sigma}}^{(M)}\cdot \hat{\vec{\sigma}}^{(A_n)}}
    \label{eq:aa_unitary}
\end{equation}
where we have defined, consistently with what done before, $\theta_{a} = g_{a}\tau_{a}$. 
The above equation is equivalent to the partial swap unitary \cite{Ciccarello21, Nielsen_Chuang_2010}
\begin{equation}
\label{eq:p_swap_unitary}
    \hat{W}_{M,A_n}(p) = \sqrt{1-p}~\mathbb{I}_{M,A_n}-i\sqrt{p}~\hat{S}_{M,A_n}
\end{equation}
with swap probability 
\begin{equation}
    p = \mathrm{sin}^2(\theta_{a}).
\end{equation}
The action of the swap operator $\hat{S}_{M,A_n}$ is such that
\begin{equation}
    \hat{S}_{M,A_n}\ket{\psi}_{M}\otimes\ket{\phi}_{A_{n}} = \ket{\phi}_{M}\otimes\ket{\psi}_{A_{n}} 
\end{equation}
for any given pair of states $\ket{\psi}_M, \ket{\phi}_{A_{n}} $\cite{Ciccarello21}. It is worth noting that we are assuming the homogeneity condition also for $g_a$ and $\tau_a$.\\
To complete the $M-A_n$ interaction scheme, the partial swap discussed above needs to be followed by a full swap so that the unitary $\hat{U}_{M,A_n}$ of the global $M-A_n$ interaction, takes the form
\begin{equation}
\label{eq:swap_p_swap_unitary}
    \hat{U}_{M,A_n}(p)=\hat{S}_{M,A_n}\hat{W}_{M,A_n}(p) = \sqrt{1-p}~\hat{S}_{M,A_n}-i\sqrt{p}~\mathbb{I}_{M,A_n}.
\end{equation}
This unitary leaves the two ancillas unchanged with probability $\mathrm{sin}^2(\theta_{a})$ and swaps them with probability $\mathrm{cos}^2(\theta_{a})$. \\

Two remarkable limiting behaviors occur for $p=0$ and $p=1$ respectively. In the former case, the interaction always swaps the states of $M$ and $A_n$, so that the subsequent $Q-M$ interaction takes place as if $Q$ interacted directly with $A_n$. This means that $A_n$ effectively replaces $M$ at every step. Thus, setting $p=0$ in each $M-A_n$ interaction recovers the Markovian (memoryless) collisional dynamics where $Q$ directly interacts with a stream of independent external chargers $\qty{A_1 \dots A_n}$. On the other hand, the $p=1$ case leaves $M$ unchanged after its interaction with $A_n$. As a consequence, the subsequent $Q-M$ interaction starts from the same state where the previous $Q-M$ interaction stopped. Thus, setting $p=1$ forces the system to repeatedly interact with $M$ without any external effect due to the chargers, so that only the memory plays a role in the system dynamics. This implies that, opposed to the $p=0$ regime, setting $p=1$ leads to fully coherent non-Markovian Rabi-like oscillations where the $n$--th step is strongly affected by the previous ones~\cite{Ciccarello21}. This justifies the fact that the swap probability $p$ is used as an operational quantifier of memory effects. Indeed, increasing $p$ progressively enhances the amount of information retained by the memory ancilla between consecutive collisions, allowing one to continuously interpolate between the memoryless ($p=0$) and the fully coherent ($p=1$) regimes.\\

It is worth noticing that the composite CM scheme discussed so far is fully equivalent, limited to the reduced dynamics of $Q$ under investigation, to a more conventional CM where no memory ancilla is present, but the chargers pairwise interact with partial swap interactions~\cite{Ciccarello21,Campbell_CM21}. However, here we have chosen the former approach because it is easier to be addressed numerically and more suitable for implementation in superconductor-based solid-state platforms, as will be discussed in Sec.~\ref{sec:exp_feas}. Moreover, it is related to more general formalisms introduced to describe structured environments such as the pseudomode approach~\cite{Pleasance20, Albarelli25} and the reaction coordinate mapping~\cite{Anto21}.  \\

Up to now we have focused on the single collisional step. In the following we will discuss the consequences of iterating the discussed composite CM dynamics.


\subsection{Composite collisional dynamics}
\label{sec:composite_CM_dynamics}
Denoting with $\hat{\rho}(0)$ the initial density matrix of the whole system, after $n$ identical steps of the composite CM dynamics, one has
\begin{equation}
    \hat{\rho}(n) = \qty(\hat{U}_{M,A_n}(p)\hat{U}_{Q,M}(\theta))\dots\qty(\hat{U}_{M,A_1}(p)\hat{U}_{Q,M}(\theta))\hat{\rho}(0)\qty(\hat{U}^\dagger_{Q,M}(\theta)\hat{U}^\dagger_{M,A_1}(p))\dots\qty(\hat{U}^\dagger_{Q,M}(\theta)\hat{U}^\dagger_{M,A_n}(p)),
\end{equation}
written in terms of the unitaries in Eq.~(\ref{eq:SM_unitary}) and~(\ref{eq:swap_p_swap_unitary}). Since the global initial state factorizes as $\hat{\rho}(0) = \dms{0}\bigotimes\hat{\eta}_M\bigotimes_{m=1}^{N} \hat{\eta}_{A_m}$, the above equation reduces to
\begin{equation}
\begin{split}
    \hat{\rho}(n) = \biggl(&\hat{U}_{M,A_n}(p)\hat{U}_{Q,M}(\theta)\dots\biggl(\hat{U}_{M,A_1}(p)\hat{U}_{Q,M}(\theta)\dms{0}\otimes\hat{\eta}_M(0)\hat{U}^\dagger_{Q,M}(\theta)\otimes\hat{\eta}_{A_1}(0)\hat{U}^\dagger_{M,A_1}(p)\biggr)\dots\\
    &\dots\hat{U}^\dagger_{Q,M}(\theta)\otimes\hat{\eta}_{A_n}(0)\hat{U}^\dagger_{M,A_n}(p)\biggr)\otimes\hat{\eta}(0)_{A_{n+1}}\otimes\dots\otimes\hat{\eta}(0)_{A_N},
\label{eq:composite_dynamics}
\end{split}
\end{equation}
that is, a sequence of applications of JC unitaries involving $Q$ and $M$, interspersed by $M-A_n$ unitaries. Let us observe that after the end of the $M-A_n$ interaction $A_n$ plays no further role and can be traced out~\cite{Ciccarello21}.\\
If the \textit{trapping condition} of Eq.~(\ref{eq:theta_ft}) holds, no $Q-M$ interaction will be able to populate the density matrix outside the size $s$ lowest-energy block~\cite{Slosser_89, Shaghaghi_MM_22,Nemeth_05,Shaghaghi_MM_23}. This is valid independently of $p$ (characterizing memory effects), which only plays a role in the $M-A_n$ interaction, and has no direct effect on the structure of $\dms{n}=\mathrm{Tr}_{M, A_{1},...,A_{n}} \hat{\rho}(n)$, which indicates the reduced density matrix of $Q$ after $n$ collision steps. Indeed, the effects of $p\neq0$ are only transferred to $Q$ through the mediation of $M$, by means of \textit{Jaynes-Cummings} interactions preserving the trapping condition. According to this, we expect the system density matrix to remain confined in the trap where it started all along the composite dynamics. Even if the fine-tuned dynamics, limiting the maximum energy accessible by the QB, does not seem the optimal choice for a QB, the size of this trap can be chosen by controlling $\theta$ (and consequently $s$), to guarantee enough states for the build up of stored energy. Moreover, the study of the Markovian limit $p=0$ has shown that this condition is preferable with respect to the non fine-tuned one, leading to better stability in terms of energy fluctuations of the QB~\cite{Shaghaghi_MM_22,Shaghaghi_MM_23}. We will provide an example of this at the beginning of Sec.~\ref{sec:results}.

\subsection{Figures of merit}
We introduce here the figures of merit used in the following to characterize the QB performances. To begin with, we consider the energy stored in the QB as a function of the number of collision steps $n$. It is given by
\begin{equation}
    E_Q(n) = \mathrm{Tr}_Q\qty{\hat{H}_Q\dms{n}}.
\end{equation}

As a measure of the QB reliability, we need also to characterize the energy fluctuations, captured by the energy variance~\cite{Friis18}
\begin{equation}
    \upvarsigma^2_Q(n) = \mathrm{Tr}_Q\qty{\hat{H}^2_Q \dms{n}}-E^2_Q(n)
\end{equation}
and the normalized parameter~\cite{Mohan26, Rinaldi26}
\begin{equation}
    \upnu_Q(n) = \frac{\upvarsigma^2_Q(n)}{E^2_Q(n)}, \quad n \geq 1.
\end{equation}
Notice that, in these relations, we have explicitly taken into account the fact that the initial state of the QB (at $n=0$) has zero energy and no fluctuations. Moreover, it is worth to mention that even if in the above expressions we have only made explicit the dependence on $n$, both the considered figures of merit also depend on other parameters that will be properly indicated and commented where needed.


\section{Results}
\label{sec:results}
In this Section, we will show the numerical results obtained for the non-Markovian charging protocol discussed above, namely for $p\neq0$. In doing so, we will focus on the JC regime, where the dynamics is ruled by the parameters $\theta$, $c$, $q$ and $p$. As stated above, the latter characterizes the non-Markovianity via partial swap of the ancillas, with $p=0$ corresponding to the memoryless limit already discussed in literature~\cite{Shaghaghi_MM_22, Shaghaghi_MM_23}. In these works, at a fixed $s$, the structure of the trapped steady state of the QB has been studied. As previously discussed, setting $p\neq0$ will not break the trap for the steady state, but it will be interesting to see how it affects the shape of the trapped density matrix in the Fock space. We will consider both fine-tuned and non fine-tuned $\theta$ values. \\

\begin{figure}[h]
\centering
     \begin{subfigure}[c]{0.43\textwidth}
         \centering
         \includegraphics[width=\textwidth]{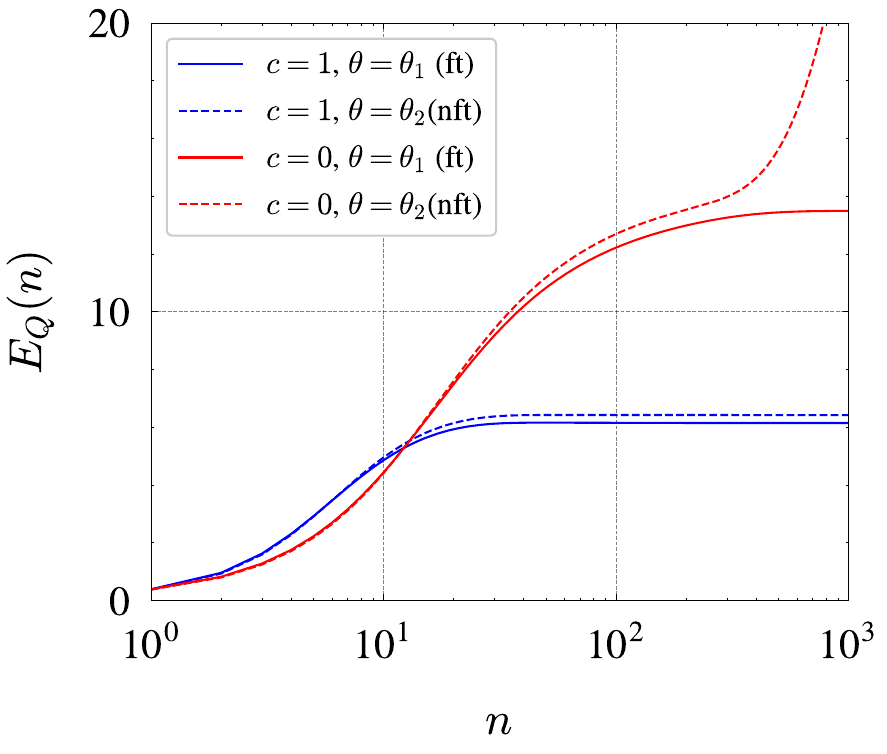}
         \caption{}
     \end{subfigure}
     \hfill 
     \begin{subfigure}[c]{0.443\textwidth}
         \centering
         \includegraphics[width=\textwidth]{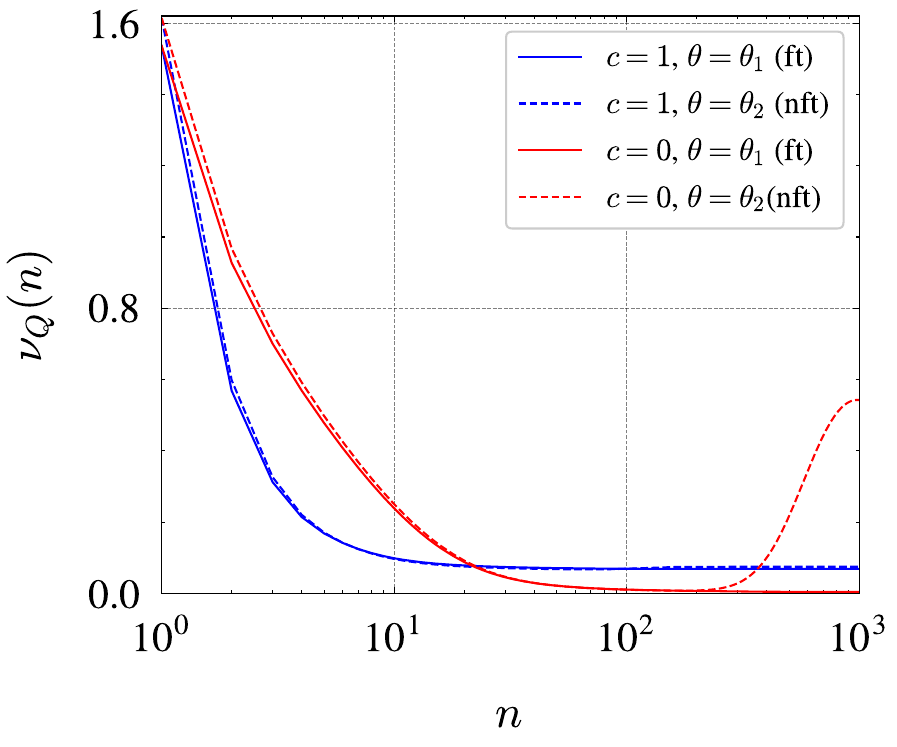}
         \caption{}
     \end{subfigure}
     
     \caption{Energy $E_Q(n)$ stored in the Micromaser QB in units of $\omega$ (a), and corresponding normalized variance $\upnu_Q(n)$ (b), as functions of the number of collisions $n$ for the Markovian limit $p=0$. Solid lines represent the fine-tuned (ft) dynamics with $\theta=\theta_1=\frac{\pi}{\sqrt{15}}$, dashed lines non fine-tuned (nft) dynamics with $\theta=\theta_2=\frac{\pi}{\sqrt{15.5}}$. Other parameters are: $s=15$ and $q=0.25$.}
     \label{fig:fig2}
\end{figure}

Before showing our results in the non-Markovian regime, let us quickly recap the situation for the Markovian ($p=0$) regime. In this case, the coherent charging protocol ($c=1$ in Eq. (\ref{eq:anc_dm})), reveals to be stable with respect to deviations from fine-tuned values of the system-ancilla parameter $\theta$. Concerning the incoherent case ($c=0$ in Eq. (\ref{eq:anc_dm})), it shows higher energy storage, but strongly affected by non-fine-tuning $\theta$, with performances worse than those of the coherent protocol in terms of stability of the stored energy~\cite{Shaghaghi_MM_23} and energy fluctuations. Examples of these behaviors are reported in Fig.~\ref{fig:fig2} confirming the coherent $c=1$ protocol to be more reliable in terms of energy stabilization and fluctuations.\\
For the above reasons, in the following we will address the coherent protocol, investigating the role of non-Markovianity, comparing our results with the Markovian case to highlight relevant differences. \\
All numerical simulations have been performed using an in-house QuTiP code~\cite{qutip5}.


\subsection{Fine-tuned JC regime}
\label{subsec:fine_tuned_JC}
\begin{figure}[h]
\centering
    \begin{subfigure}[c]{0.3\textwidth}
        \centering
        \includegraphics[width=\textwidth]{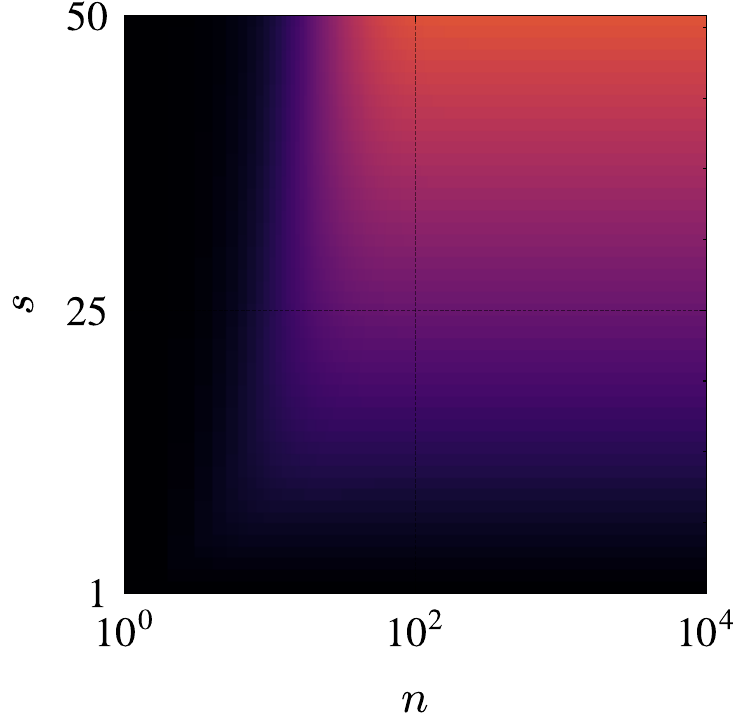}
        \caption{$p=0$}
    \end{subfigure}
    \hfill 
    \begin{subfigure}[c]{0.3\textwidth}
        \centering
        \includegraphics[width=\textwidth]{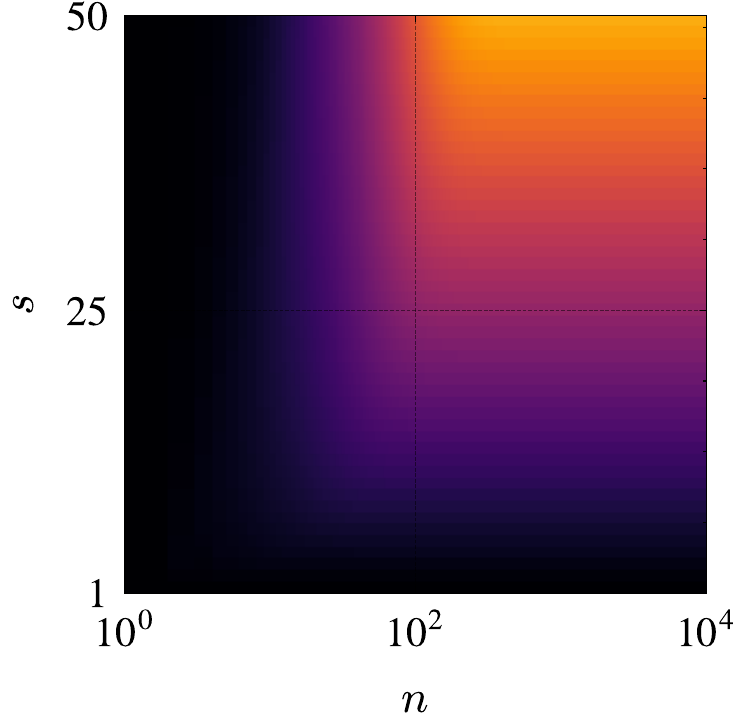}
        \caption{$p=0.25$}
    \end{subfigure}
    \hfill 
    \begin{subfigure}[c]{0.3\textwidth}
        \centering
        \includegraphics[width=\textwidth]{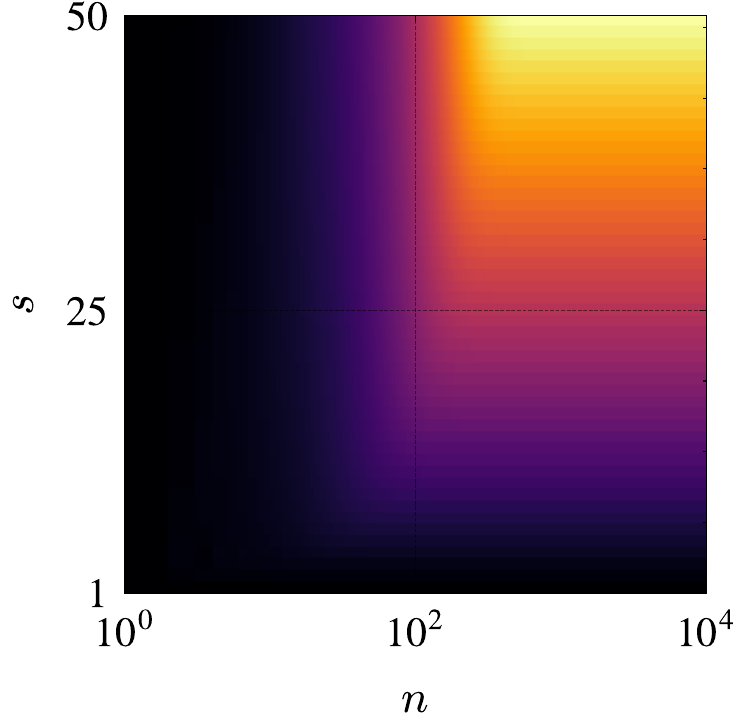}
        \caption{$p=0.5$}
    \end{subfigure}
    \hfill
    \begin{subfigure}[c]{0.07\textwidth}
        \centering
        \includegraphics[width=\textwidth]{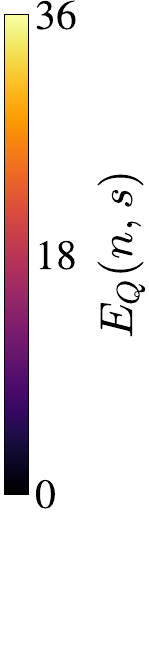}
    \end{subfigure}
     
     \caption{Energy $E_Q(n,s)$ stored in the Micromaser QB, in units of $\omega$, as a function of the number of collisions $n$ and for different trap sizes $s$. Comparison between the Markovian (a) and non-Markovian (b)-(c) charging protocols. Other parameters are: $c=1$ and $q=0.25$.}
     \label{fig:fig3}
\end{figure}

\begin{figure}[h]
    \includegraphics[scale = 0.45]{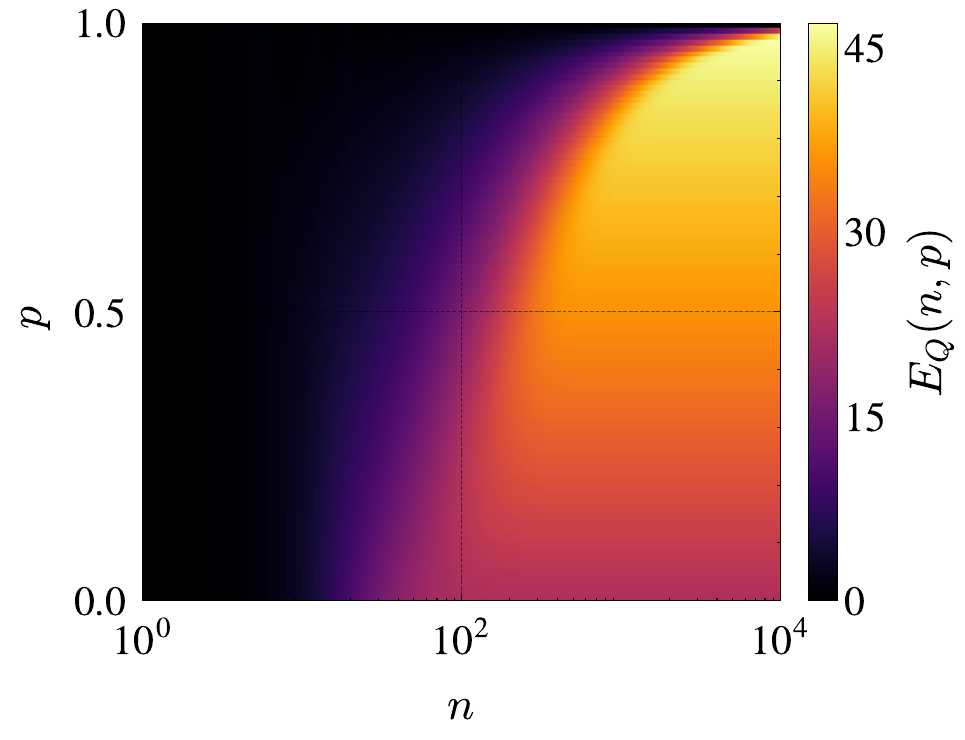}  
    \centering
    \caption{Energy $E_Q(n,p)$ stored in the Micromaser QB, in units of $\omega$, as a function of the number of collisions $n$ and the swap parameter $p$ for a trap size value $s=50$ (corresponding to the greatest value reported in Fig. \ref{fig:fig3}). Other parameters are $c=1$ and $q=0.25$.}
     \label{fig:fig4}
\end{figure}

We start with a discussion of the fine-tuned dynamics ($\theta = \theta_{ft}=\frac{\pi}{\sqrt{s}}$). Fig.~\ref{fig:fig3} shows the density plot of the stored energy as a function of $n$ and $s$, for three representative values of swap probability $p$.\\ 
All panels ((a)-(c)) show common features: for all the values of $s$, the stored energy increases with $n$, reaching a plateau similar to the ones observed in Fig.~\ref{fig:fig2}. Furthermore, increasing $s$ increases both the stored energy and the time needed to stabilize to the asymptotic value, since the energy plateau moves to the right. \\
Comparing the results obtained for the Markovian case $p=0$ (in panel (a)) with the non-Markovian ones, namely $p=0.25$ (panel (b)) and $p=0.5$ (panel (c)), we observe that memory effects yield higher values of the asymptotic stored energy (brighter regions), reached at the price of longer charging times (greater $n$ needed for reaching the excited plateau). Therefore, this enhanced charging can be of practical advantage only if such a steady state is obtained within time scales compatible with the platforms considered for realization of the charging protocol. To better characterize this point, we now fix a given value of $s$ (a fixed fine-tuned $\theta_{ft}$ according to Eq.~(\ref{eq:theta_ft})) and study the stored energy as a function of both the number of collisions $n$ and the swap probability (see Fig.~\ref{fig:fig4}). Here, a clear trade-off emerges: together with the enhancement of the stored energy, the charging time increases when moving towards $p=1$. Thus, regimes of intermediate values of $p$ could represent a good compromise between having a relevant stored energy and a short enough charging time. At the edge value for the swap probability, namely $p=1$, the excited steady state is never reached. 

Indeed, as discussed in Sec.~\ref{subsubsec:M_A_interactions}, $p=1$ corresponds to the situation where the QB interacts only with the memory ancilla, giving rise to a coherent oscillating dynamics with no steady state\footnote{This limit is of no practical interest for a QB charging, since the maximum amount of energy transferable to the QB would correspond only to the energy initially stored into the memory $M$, namely $E_M(0) = (1-q) $ in units $\omega$, explaining the dark stripe at $p=1$ in Fig.~\ref{fig:fig4}.}. \\

Since the composite non-Markovian dynamics as described in Eq.~(\ref{eq:composite_dynamics}) develops as a sequence of maps applied iteratively to the $Q-M$ joint state, all dynamically equivalent due to the homogeneity assumption for the ancillas, an interesting analysis is the one concerning the steady state of the battery subject to such process, obtained by following the procedure discussed in Appendix~\ref{Steady}. This allows us to understand where the charging drives the battery at long times $\left(n \rightarrow \infty \right)$. At this purpose, we analyze  the energy stored in the steady state $E^{st}_Q=E_Q(n\rightarrow+\infty)$ and its normalized variance $\upnu^{st}_Q=\upnu_Q(n\rightarrow +\infty)$ both as functions of $p$ and $s$. \\
\begin{figure}[h]
\centering
     \begin{subfigure}[c]{0.43\textwidth}
         \centering
         \includegraphics[width=\textwidth]{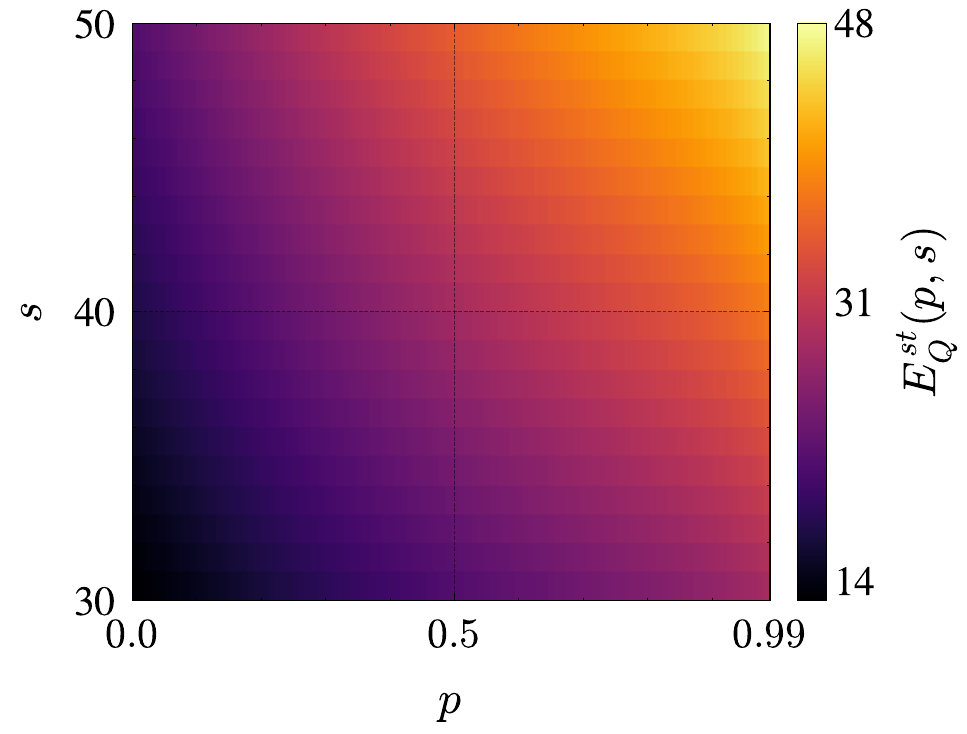}
         \caption{}
     \end{subfigure}
     \hfill 
     \begin{subfigure}[c]{0.48\textwidth}
         \centering
         \includegraphics[width=\textwidth]{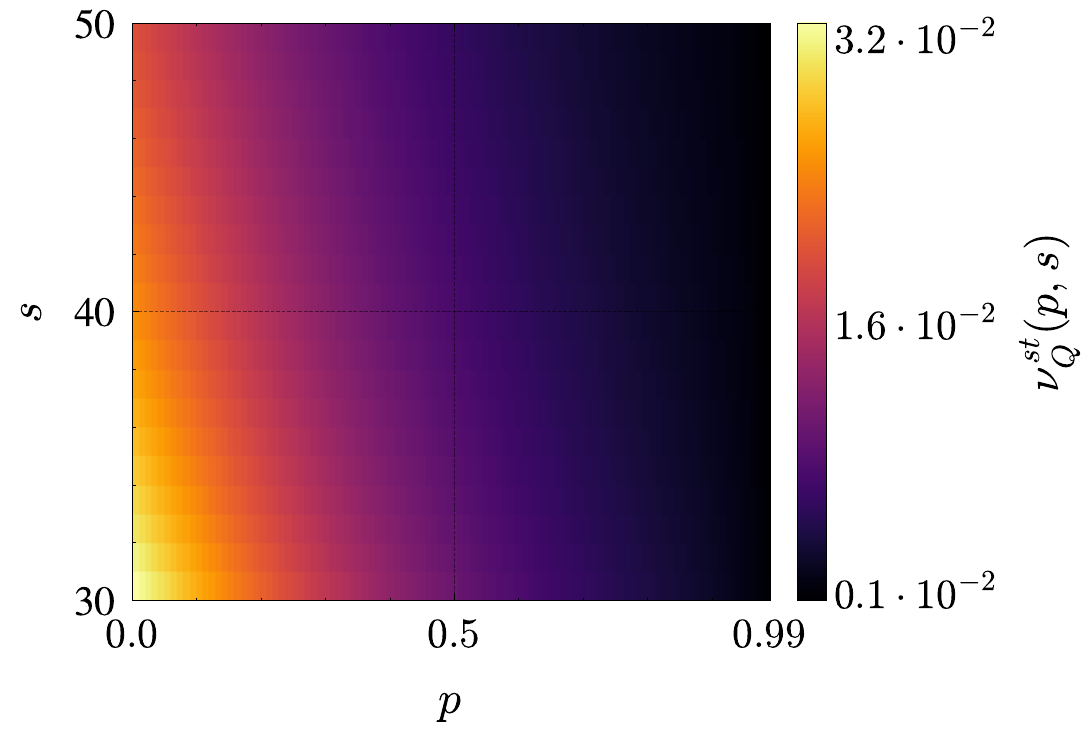}
         \caption{}
     \end{subfigure}
     
     \caption{Steady state energy $E^{st}_Q$, in units of $\omega$, (a) and corresponding normalized variance $\upnu^{st}_Q$ (b) as functions of the swap probability $p$ and of the trap size $s$. Other parameters are: $c=1$ and $q=0.25$.}
     \label{fig:fig5}
\end{figure}
Fig.~\ref{fig:fig5} clearly shows that transitioning from the Markovian regime $p=0$ up to strongly non-Markovian regimes with $p=0.99$ leads to a relevant increase in $E^{st}_Q$ (panel (a)), simultaneously suppressing energy fluctuations $\upnu^{st}_Q$ (panel (b)).\\
As discussed above, the approach used to determine the steady state of the system fails for $p=1$, which is the reason for which we have focused in the range $p\in[0,0.99]$ in Fig.~\ref{fig:fig5}.\\

Thus, intermediate regimes of non-Markovianity $0<p<1$  are characterized by enhanced steady state stored energy and suppressed fluctuations with respect to the Markovian case.\\

\begin{figure}[h]
\centering
     \begin{subfigure}[c]{0.29\textwidth}
         \centering        \includegraphics[width=\textwidth]{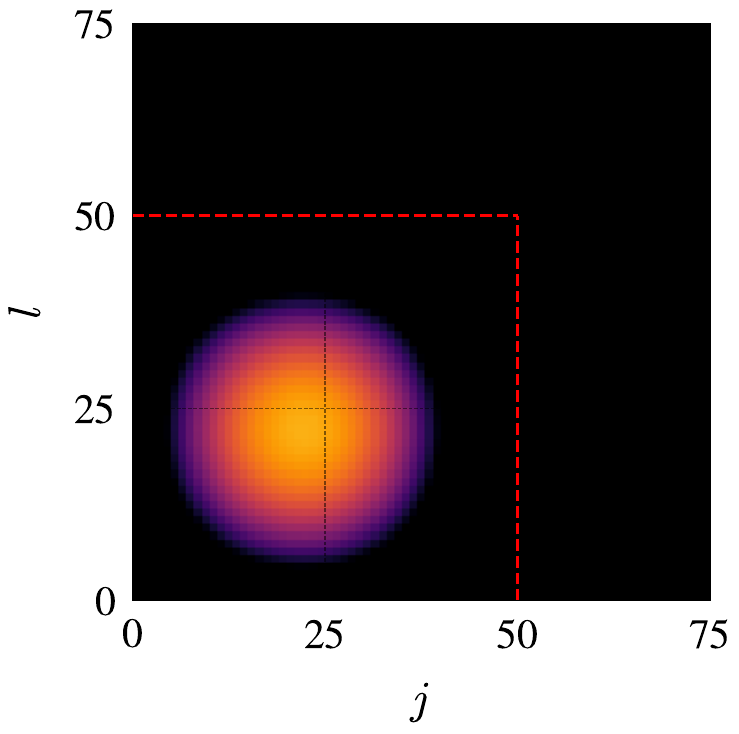}
         \caption{}
     \end{subfigure}
     \hfill 
     \begin{subfigure}[c]{0.29\textwidth}
         \centering
         \includegraphics[width=\textwidth]{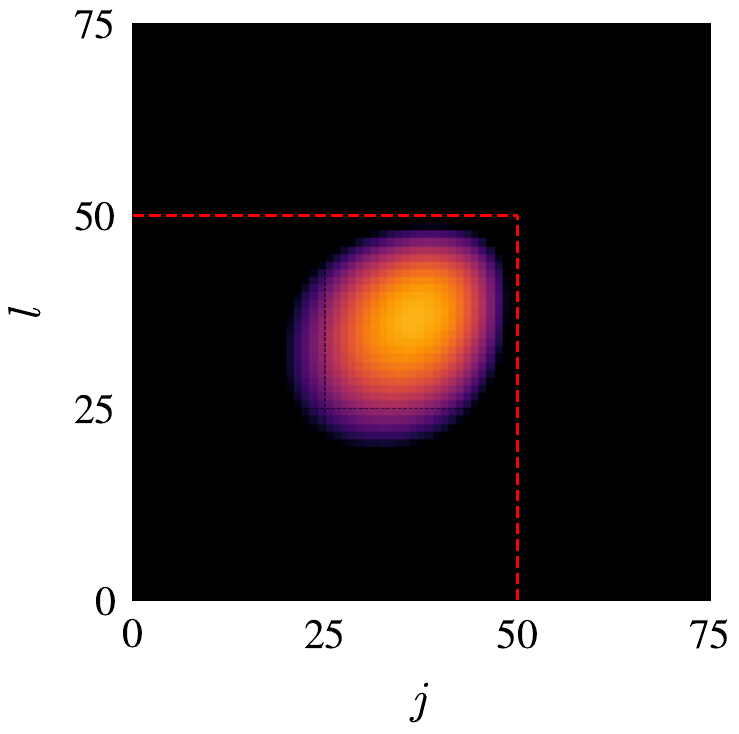}
         \caption{}
     \end{subfigure}
     \hfill 
     \begin{subfigure}[c]{0.29\textwidth}
         \centering
         \includegraphics[width=\textwidth]{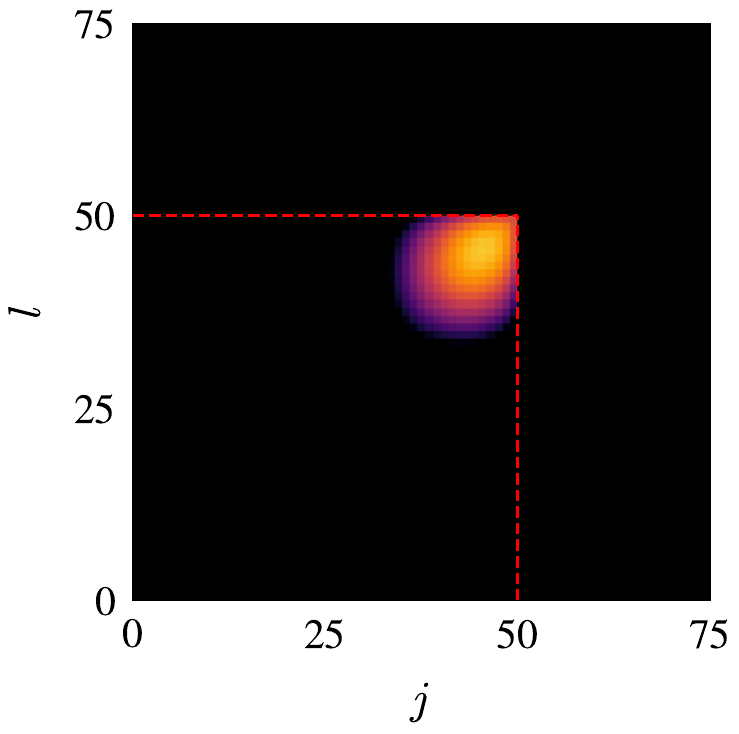}
         \caption{}
     \end{subfigure}
     \hfill
    \begin{subfigure}[c]{0.1\textwidth}
        \centering
        \includegraphics[width=\textwidth]{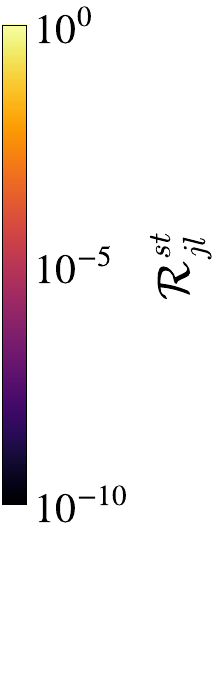}
    \end{subfigure}
     
     \caption{Squared modulus $\mathcal{R}^{st}_{jl}$ of the entries of the QB density matrix for the steady state in the Markovian protocol $p=0$ (a) and a non-Markovian protocols with $p=0.5$ (b) and $p=0.9$ (c). Other parameters are: $s=50$, $c=1$ and $q=0.25$.}
     \label{fig:fig6}
\end{figure}

We now turn to a physical explanation of the origin of the larger stored energy and suppressed fluctuations in the non-Markovian regime. In order to do so, we will inspect the squared modulus of the entries of the steady state density matrix of the battery $\mathcal{R}^{st}_{jl}=|\langle j|\dms{n\rightarrow +\infty}|l\rangle|^{2}$ ($j$ and $l$ indicating Fock states of the harmonic oscillator QB). Fig.~\ref{fig:fig6} compares the Markovian case (panel (a)) with two non-Markovian conditions (panels (b)-(c)) for a fixed value of the fine-tuned coupling, namely a fixed trap size $s=50$ for all panels. As expected, memory effects introduced with $p\neq0$ transform the steady state density matrix without breaking the limit imposed by the trap (red dashed line). \\
By increasing $p$, the relevantly populated region of the density matrix simultaneously shrinks and moves towards the upper bound of the trap. The progressive localization of the steady-state distribution close to the upper part of the trap increases the energy of the state and reduces the statistical spread of the populated Fock states leading to a suppression of the energy fluctuations. This indicates that the memory ancilla acts actively as a temporary energy reservoir, creating correlations between successive charging events and forcing the accumulation of population close to the upper boundary of the trapping chamber. Thus, the non-Markovian protocols are able to better exploit the trapping dynamics, moving the steady states towards higher energies in a more stable configuration. \\

\begin{figure}[h]
\centering
     \begin{subfigure}[c]{0.29\textwidth}
         \centering        \includegraphics[width=\textwidth]{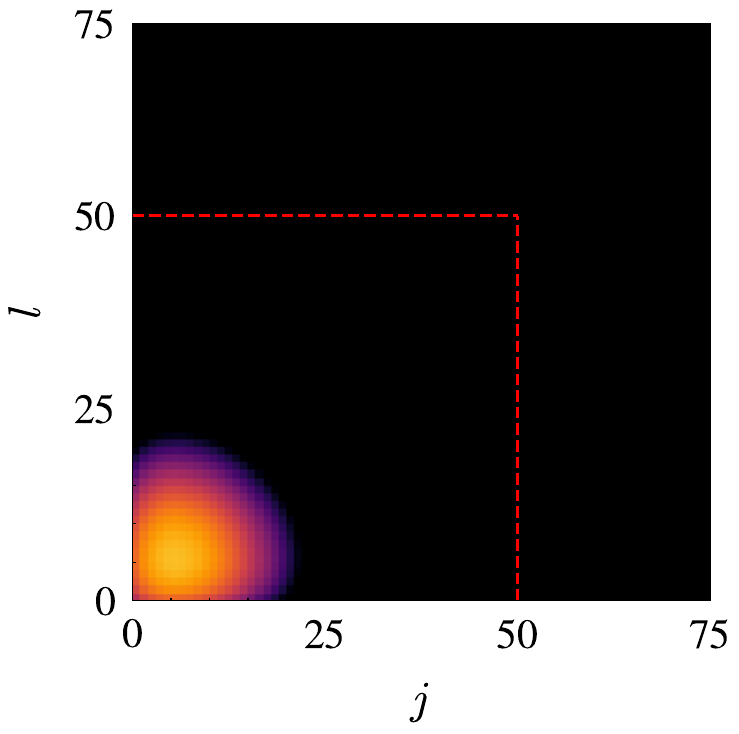}
         \caption{}
     \end{subfigure}
     \hfill 
     \begin{subfigure}[c]{0.29\textwidth}
         \centering
         \includegraphics[width=\textwidth]{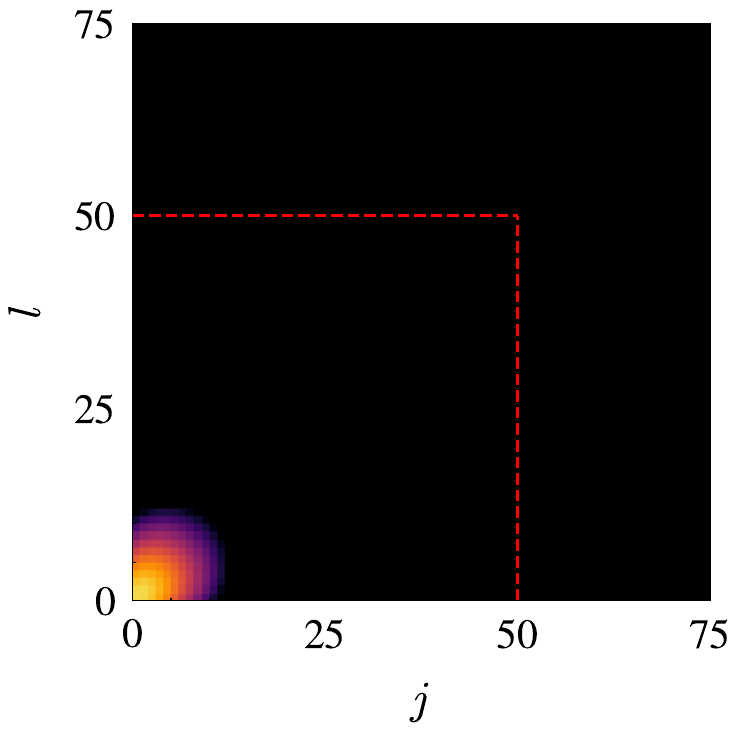}
         \caption{}
     \end{subfigure}
     \hfill 
     \begin{subfigure}[c]{0.29\textwidth}
         \centering
         \includegraphics[width=\textwidth]{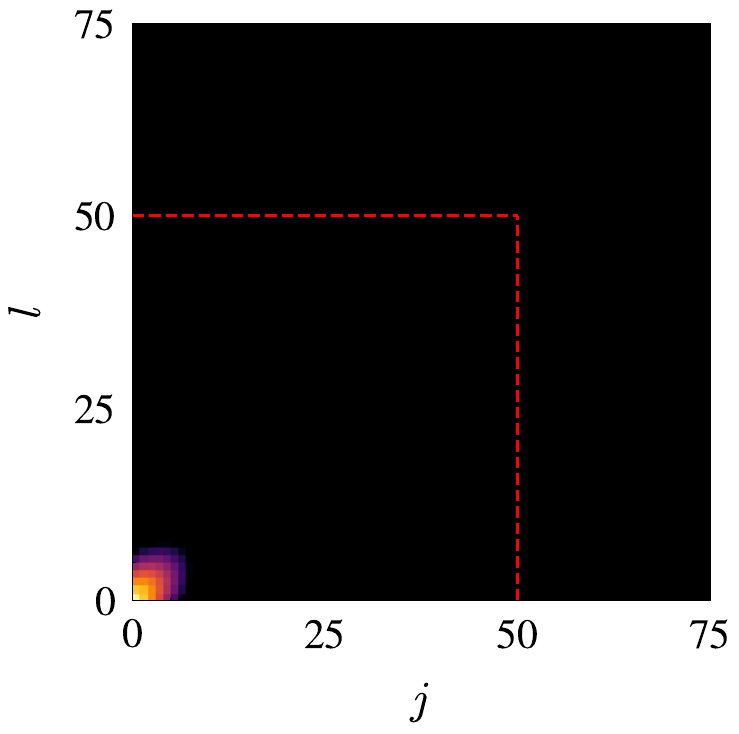}
         \caption{}
     \end{subfigure}
     \hfill
    \begin{subfigure}[c]{0.1\textwidth}
        \centering
        \includegraphics[width=\textwidth]{images/fig6cbar.pdf}
    \end{subfigure}
     
     \caption{Squared modulus $\mathcal{R}^{st}_{jl}$ of the entries of the QB density matrix for the steady state in the Markovian protocol $p=0$ (a) and a non-Markovian protocols with $p=0.5$ (b) and $p=0.9$ (c). Other parameters are: $s=50$, $c=1$ and $q=0.75$.}
     \label{fig:fig7}
\end{figure}

It is important to stress that the observed results are not limited to the considered value of $q=0.25$. Indeed, qualitatively analogous behavior also emerges for other values of $q<0.5$. \\
Radically different is the situation for the complementary case $q>0.5$ (see Fig. \ref{fig:fig7}), a regime where ancillas populations are not inverted. Already in the Markovian protocol this case appears less convenient for a QB setup, since it corresponds to less energetic ancillas. The role played by non-Markovianity here further amplifies this situation. Indeed, the steady state structure, still trapped as expected, is progressively brought towards the lower part of the trap by increasing $p$. This demonstrates that the advantages associated to the memory effects are crucially related to the energy initially stored in the chargers. 

\subsection{Robustness with respect to non-fine-tuning}
\begin{figure}[h]
\centering
     \begin{subfigure}[c]{0.48\textwidth}
         \centering
         \includegraphics[width=\textwidth]{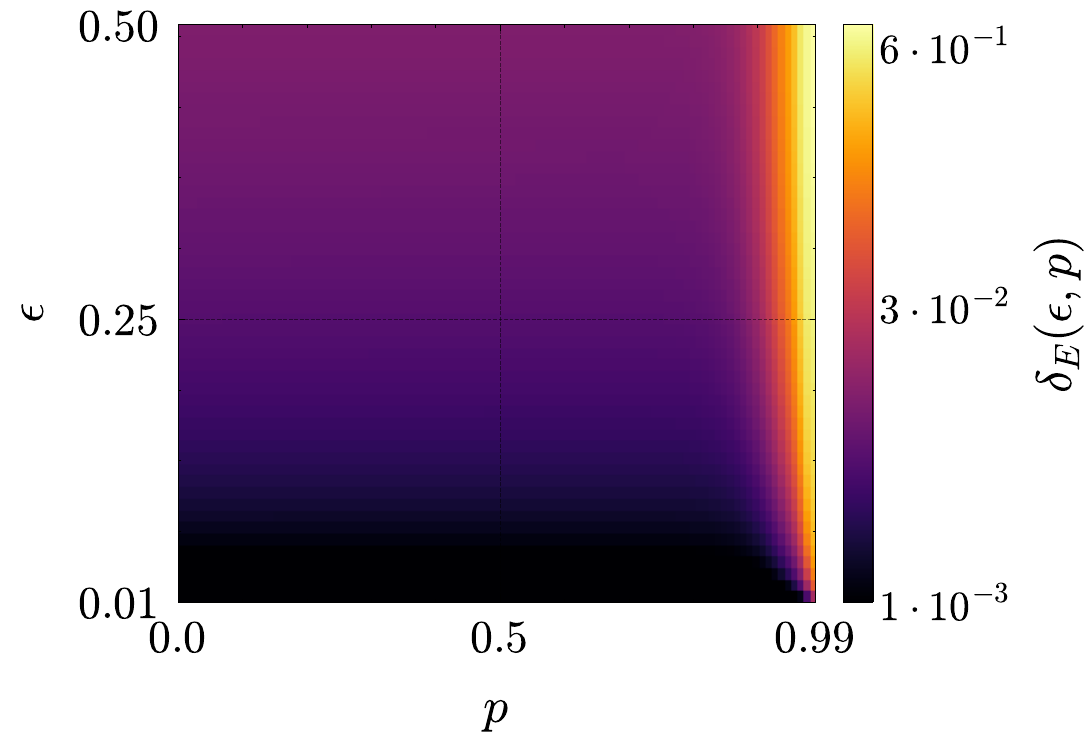}
         \caption{}
     \end{subfigure}
     \hfill 
     \begin{subfigure}[c]{0.48\textwidth}
         \centering
         \includegraphics[width=\textwidth]{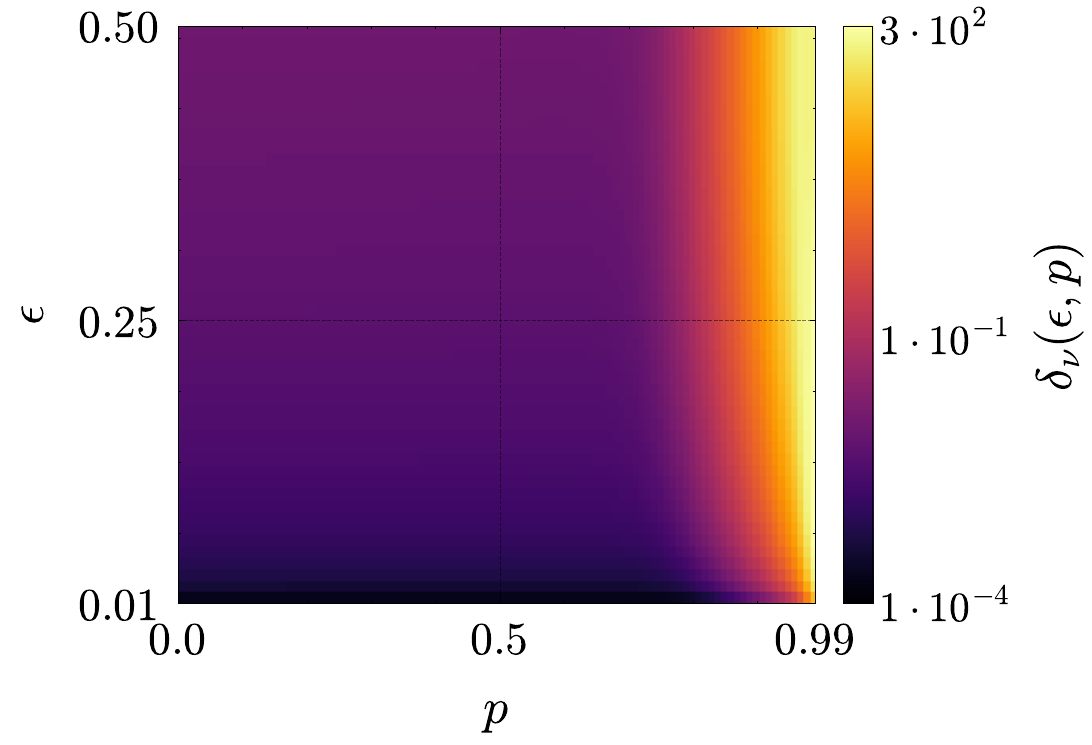}
         \caption{}
     \end{subfigure}
     
     \caption{Relative variations $\delta_E(\epsilon,p)$ for the steady state stored energy (a) and $\delta_\upnu(\epsilon,p)$ for its variance (b) as functions of the swap probability $p$ and the detuning $\epsilon$.  Other parameters are: $s=50$, $c=1$ and $q=0.25$.}
     \label{fig:fig8}
\end{figure}

As mentioned before, in the Markovian case it has been shown that deviations with respect to the fine-tuned condition $\theta=\theta_{ft}$ can create a breach in the trap where the fine-tuned dynamics is confined~\cite{Shaghaghi_MM_22, Shaghaghi_MM_23}. As a consequence, charging protocols pushing the steady state to the limit of the trap show relevant leakages towards higher energy states and are affected by an increased instability. This is the case observed for Markovian incoherent charging protocols~\cite{Shaghaghi_MM_23} (see Fig.~\ref{fig:fig2}). Since, as shown in Fig.~\ref{fig:fig6}(b-c), a non-Markovian coherent charging protocol is even more efficient in pushing the steady state towards the limit of the trap, it is worth discussing its stability away from the fine-tuned case.\\
Non-fine-tuned values of the coupling can be parametrized as 
\begin{equation}
    \theta = \frac{\pi}{\sqrt{s+\epsilon}}, \quad \epsilon \neq 0 
\end{equation}
in terms of the detuning parameter $\epsilon \in \qty[-0.5,0.5]$. Values of $\epsilon$ outside this range can be mapped in the above condition simply by changing the trap size $s$. In order to directly compare the fine-tuned and non-fine-tuned dynamics, we consider the steady state stored energy and characterize its relative variation due to non-fine-tuning, namely
\begin{equation}
    \delta_E(\epsilon,p) = \frac{E^{st}_Q(\epsilon,p)-E_Q^{st}(0,p)}{E_Q^{st}(0,p)}.
\end{equation}
Analogously, we define an estimator for the deviations in the fluctuations as 
\begin{equation}
    \delta_{\upnu}(\epsilon,p) = \frac{\upnu^{st}_Q(\epsilon,p)-\upnu_Q^{st}(0,p)}{\upnu_Q^{st}(0,p)}.
\end{equation}
Numerical results for these quantities are shown in Fig.~\ref{fig:fig8}. 

\begin{figure}[h]
\centering
     \begin{subfigure}[c]{0.29\textwidth}
         \centering
         \includegraphics[width=\textwidth]{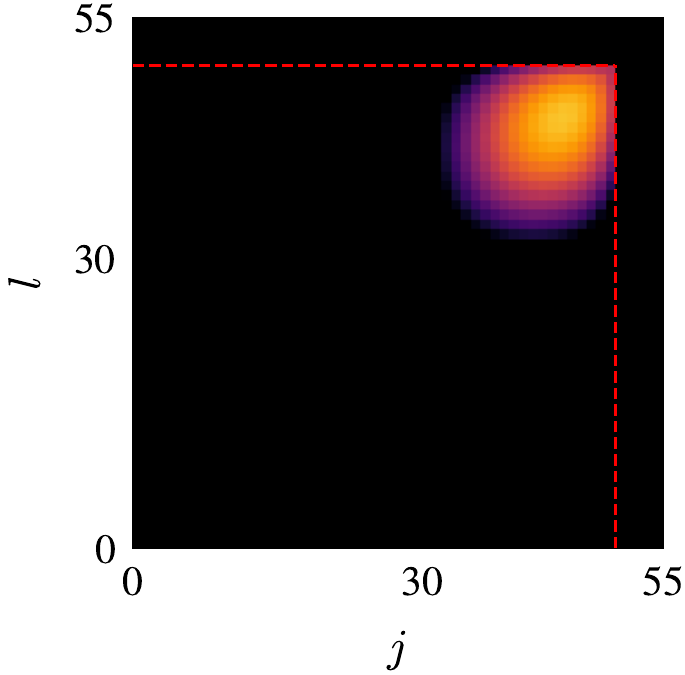}
         \caption{}
     \end{subfigure}
     \hspace{0.1cm}
     \begin{subfigure}[c]{0.1\textwidth}
        \centering
        \includegraphics[width=\textwidth]{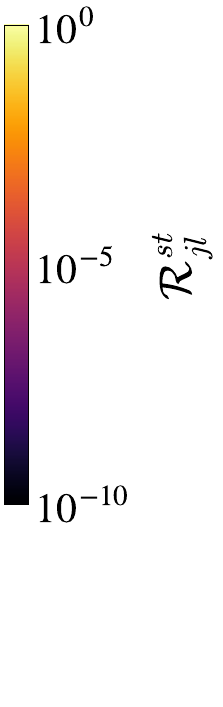}
    \end{subfigure}
    \hfill
     \begin{subfigure}[c]{0.29\textwidth}
         \centering
         \includegraphics[width=\textwidth]{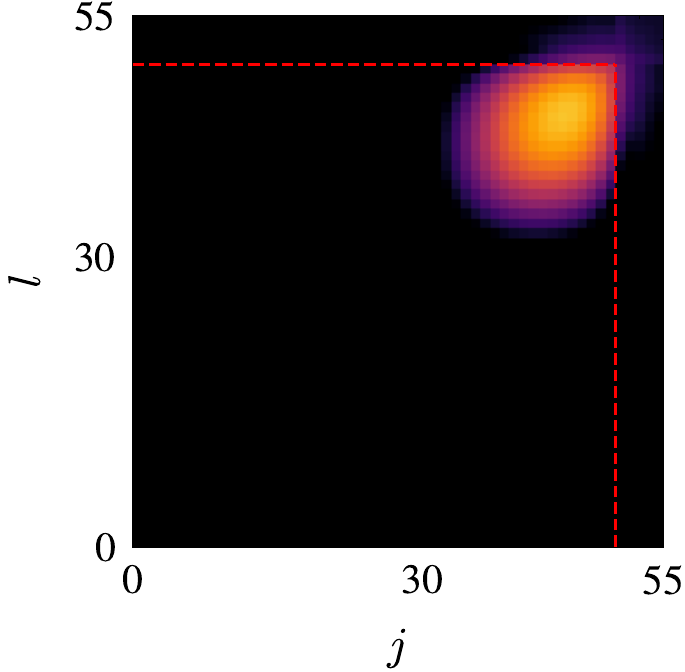}
         \caption{}
     \end{subfigure}
     \hspace{0.1cm}
     \begin{subfigure}[c]{0.1\textwidth}
        \centering
        \includegraphics[width=\textwidth]{images/fig9cbar.pdf}
    \end{subfigure}
     \caption{Squared modulus $\mathcal{R}^{st}_{jl}$ of the entries of the QB density matrix for the steady state in the non Markovian protocol for a fine-tuned case $\epsilon=0$ (a) and a non fine-tuned case with $\epsilon=0.25$. where a leakage outside the trap emerges (b). Other parameters are: $s=50$, $p=0.86$, $c=1$ and $q=0.25$.}
     \label{fig:fig9}
\end{figure}

For what it concerns the steady state stored energy (panel (a)), two different behavior can be identified: for $p\lesssim 0.9$, increasing $\epsilon$ leads to deviations of the stored energy of the order of a few percent at most; for $p\gtrsim 0.9$, the deviations abruptly increase, reaching values up to $60\%$. Something similar happens also for the variations of the steady state energy fluctuations, shown in panel (b). Here, however, $\delta_\upnu$ is more sensitive to $\epsilon $ and it exceeds $10\%$ when $p \gtrsim0.75$, reaching very high values in the strongly non-Markovian region $p \gtrsim0.9$. \\
To see that this behavior is linked to the breaking of the trapping condition, Fig.~\ref{fig:fig9} shows examples of steady states obtained for $s=50$ and $p=0.86$, namely $p$ in the region where the trap breaking starts. Here, we compare the fine-tuned steady state $\epsilon=0$ (panel (a)) with a non fine-tuned one with intermediate value for the detuning parameter $\epsilon=0.25$ (panel (b)). The latter shows the process leading to deviations of Fig.~\ref{fig:fig8}: non fine-tuning opens a breach in the trap and the battery steady state shows a leakage towards the outer states.\\
However, as shown in Sec.~\ref{subsec:fine_tuned_JC}, the less stable region is also characterized by longer charging times, requiring thousands of collisions to saturate the steady state stored energy. Thus, stability with respect to non-fine-tuning strengthens the idea that the optimal working point for the non-Markovian charging protocol lies in a regime of intermediate values of $p$. 


\section{Remarks on experimental feasibility}
\label{sec:exp_feas}
We comment now about possible experimental implementations of the non-Markovian Micromaser QB discussed so far. The most natural candidates for implementing a memoryless Micromaser would be cavity quantum electrodynamics setups where a stream of suitably initialized atoms are injected one by one into a high quality superconducting cavity~\cite{Deleglise2008,Haroche13}. \\
The major issue in adding non-Markovianity in this scheme is posed by the possibility of actually engineering and controlling the proper atom-atom exchange interaction. Motivated by this fact, we propose here a solid-state implementation based on superconducting circuits. Indeed, due to the astonishing technological progress that has occurred in the last decades both at the level of fabrication and electronic control, these platforms currently allow to engineer the required interactions in integrated circuits~\cite{Krantz19,Blais_04,Xiang_13,Gu_17}.

\subsection{JC interaction between QB and memory ancilla}
In circuit quantum electrodynamics setups, the role of the single-mode cavity can be played by a high quality superconducting resonator, namely an LC circuit, which can be quantized in terms of the single-mode harmonic oscillator representing the QB. In addition, TLSs can be realized by means of superconducting qubits such as transmons~\cite{Goppl_2008,Krantz19,Blais_04, Xiang_13, Gu_17}. 
Both devices can be operated in the GHz range~\cite{Xiang_13, Gu_17, Krantz19, Goppl_2008}.\\

The JC interaction discussed above can be realized through capacitive coupling between the resonator and the ancillary qubit~\cite{Blais_04,Krantz19,Ciani_2024,Xiang_13}. Since our theoretical results assume no external dissipation for the oscillator and the qubit, together with the validity of the RWA, one must guarantee~\cite{Kockum19,Gu_17}
\begin{equation}
\label{eq:coupl_regimes}
    \text{max}\qty{\gamma, \kappa} \ll g \ll \omega,
\end{equation}
where $\gamma, \kappa$ represent respectively the qubit and resonator decay rates. In state-of-the-art quantum circuits these rates typically range in the $10-100$ KHz interval~\cite{Krantz19, Goppl_2008}. For what it concerns the upper bound, the validity of the RWA is assured to be safe by assuming $g/\omega \lesssim 0.05$. Given a typical range of $\omega\approx 5-10$ GHz for both superconducting cavities and qubit~\cite{Xiang_13, Gu_17,Krantz19}, one would need a capacitive coupling of the order of $g \lesssim 250$ MHz, which is way higher than the decay rates and well within reachable experimental conditions~\cite{Blais_04,Krantz19, Xiang_13}. Moreover, the typical duration of accessible quantum operations on these devices is in the order of a few tens of nanoseconds, covering all values of $\theta=g\tau$ considered in our simulations. 


\subsection{Ancillary interactions}
The chargers can be initialized in a coherent superposition of ground and excited states via a capacitive coupling with an external microwave control~\cite{Krantz19,Ciani_2024}. As shown above, the required unitary for the $M-A_n$ interaction is a partial swap followed by a full swap. The latter can be engineered through a capacitive coupling between the two TLSs, usually in the range $g_{fs}\approx 10-50$ MHz, leading to full swap times $\tau_{fs}=\frac{\pi}{2g}\approx 30-150 $ ns~\cite{Krantz19, Ciani_2024}.\\
For what it concerns the partial swap unitary $\hat{W}_{M, A_{n}}$, one need to engineer the two qubit interaction Hamiltonian of Eq.~(\ref{eq:aa_int_pot}). It contains both a transverse $\qty(\sigma_x\sigma_x + \sigma_y\sigma_y)$ and longitudinal $\qty(\sigma_z\sigma_z)$ contribution, and a direct capacitive coupling is not enough to implement both these terms~\cite{Krantz19}. However, recent developments in Floquet qubits architectures have shown the possibility of engineering the required exchange interaction between two transmon qubits by means of three microwave pulses applied to them via transmission lines~\cite{Nguyen_24}. In this scheme the reported pulse durations for a full swap operation $(p=1)$ is an order of magnitude longer than the one for $Q-M$ interactions ($\tau_{a}(p=1) \approx 100$ ns), but still compatible with our analysis. It is worth to comment that, due to the fact that $\tau_{a} \sim \mathrm{arcsin}\qty(\sqrt{p})$, intermediate regimes for the swap probability $p$, which appear to be advantageous in terms of energy storage and stability of the QB, could be obtained in shorter times. The possibility of improving this architecture and achieving faster gates for this interaction has also been considered~\cite{Nguyen_24}, possibly leading to a further speed-up.


\subsection{Overall platform and timing}
\begin{figure}[h]
    \includegraphics[scale = 0.45]{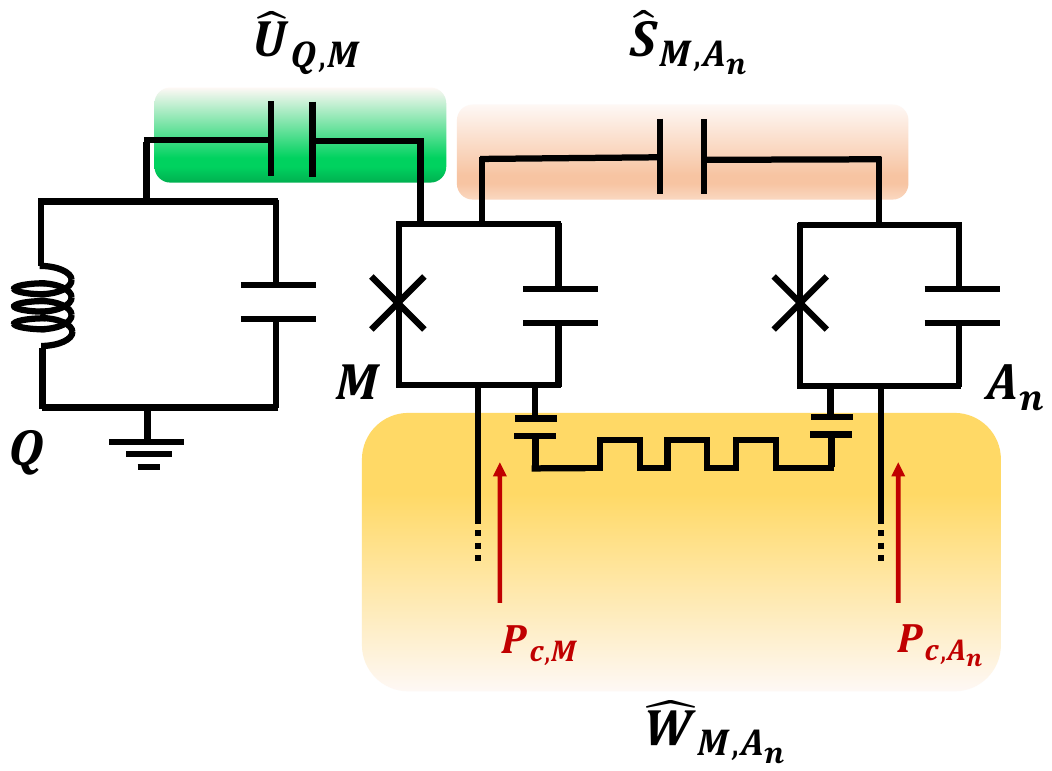}  
    \centering
    \caption{Scheme of the circuit proposed for realizing the non-Markovian Micromaser QB. The memory ($M$) and chargers ($A_{n}$) are realized by means of superconducting transmon circuits in the qubit limit, represented by boxes where the $X$ symbol denote the SQUID, which is shunted by a capacitance in parallel. The JC interaction is realized through capacitive coupling of the resonator QB with the memory ancilla (green box). The $M-A_n$ interaction is given by the partial swap followed by a full swap implemented by the yellow and red circuit elements respectively. The first is implemented via a mediating stripline resonator (squared line), and by applying controlled external pulses $P_{c,M}, P_{c,A_n}$ through transmission lines coupled to them~\cite{Nguyen_24}. The second one is realized by means of a capacitive coupling.}
     \label{fig:fig10}
\end{figure}

A minimal scheme for the platform proposed for the composite CM of Sec.~\ref{subsec:Mem_assisted_MM} is shown in Fig.~\ref{fig:fig10}.\\
It is composed by a superconducting resonator (QB) and two transmon qubits, one employed as memory $\qty(M)$ and the other, reinitialized at each step, mimiking the sequence of chargers  $\qty(A_n)$. As discussed in the main text the single charging step involves a $Q-M$ interaction followed by a $M-A_n$ one, which in turn would be composed by a partial swap followed by a full swap. According to the above estimates, it is reasonable to assume an overall duration $\tau_{step} = \tau+\tau_{fs}+\tau_{a} \approx 200-300$ ns for state-of-the-art circuits of this kind.\\

After the first of such steps ($n=1$), the process can be iterated as follows: the QB and the memory ancilla are again coupled; the charger is reinitialized to the initial ancillary state desired; the memory and the charger are again coupled. The time required to reinitialize the charger is of the order of a few nanoseconds~\cite{Xiang_13,Werninghaus_21}, so that the re-initialization can be carried out during the $Q-M$ interaction. Due to long relaxation times of the transmon platform considered for the ancillas~\cite{Koch_07,Krantz19, Blais_04, Xiang_13}, quantum coherences are able to survive during the entire step. Furthermore, the decay times of the radiation stored in high quality-factor superconducting resonators can reach values on the order of $100 $ $\mu$s. Assuming the upper bound of the estimate of the single step duration, namely $\tau_{step} \approx 300$ ns, this timescale allows to perform beyond $n = 300$ collisions and reach the advantage provided by an intermediate value of $p$ (see Fig.~\ref{fig:fig4}).


\section{Conclusions}
We have investigated the charging of a Micromaser quantum battery subject to a controlled non-Markovian charging. Following a composite collisional approach, memory has been introduced through a two-level memory ancilla mediating the interaction between a resonant cavity mode and a stream of charging qubits. This setup allows to interpolate between Markovian and non-Markovian charging regimes. 

Our numerical analysis demonstrates that memory effects can substantially improve the performance of this device. In particular, non-Markovian effects enhance the amount of energy stored in the cavity while simultaneously reducing energy fluctuations. This identifies memory as a useful resource for quantum energy-storage protocols, enabling charging processes that are not only more effective but also more reliable, where intermediate values of the partial swap probability $p$ provide the best compromise between large stored energy and controlled charging times, revealing a trade-off between speed and quality of the charging process. 

Going beyond the specific Micromaser implementation considered here, our results highlight the potential of engineered environmental memory as a tool for the optimization of quantum batteries and, more generally, of quantum thermodynamic devices. Since the proposed architecture can be implemented in currently available superconducting circuits architectures, our work opens the way to the experimental investigation of memory-enhanced charging protocols and their exploitation in future quantum-energy applications.

\ack{Sample text inserted for demonstration.}


\roles{Sample text inserted for demonstration.}

\data{Data have been produced by means of a in-house built QuTip code. They will be provided upon reasonable request.}


\appendix

\section{Trapping chambers for the Jaynes-Cummings dynamics} \label{Trapping}

In order to clarify the emergence of the trapping behavior associated to the JC dynamics we focus on the Markovian case. Here, it is possible to derive an analytical expression for the QB in the form of the quantum map \cite{Shaghaghi_MM_23, Ciccarello21}
\begin{equation}
\label{eq:Markov_qmap}
\begin{split}
\dms{n} = (1-q)&\left\{ \hat{B}\dms{n-1}\hat{B} + \hat{a}^\dagger \hat{S}\dms{n-1}\hat{S}\hat{a} \right\} \\
+ q&\left\{ \hat{C}\dms{n-1}\hat{C} + \hat{S}\hat{a}\dms{n-1}\hat{a}^\dagger \hat{S} \right\} \\
+ ic\sqrt{q(1-q)}&\left\{ \hat{C}\dms{n-1}\hat{S}\hat{a} - \hat{a}^\dagger \hat{S}\dms{n-1}\hat{C} + \right. \\
&\left. + \hat{B}\dms{n-1}\hat{a}^\dagger \hat{S} - \hat{S}\hat{a}\dms{n-1}\hat{B} \right\}
\end{split}
\end{equation}
with
\begin{align}
    \hat{B} &= \mathrm{Cos}\biggl(\theta \sqrt{\hat{N}+\mathbb{I}_S}\biggr) \\
    \hat{C} &= \mathrm{Cos}\bigl(\theta\sqrt{\hat{N}}\bigr)\\
    \hat{S} &= \frac{\mathrm{Sin}\bigl(\theta  \sqrt{\hat{N}+\mathbb{I}_S}\bigr)}{\sqrt{\hat{N}+\mathbb{I}_S}}\\
    \hat{N} &= \hat{a}^\dagger \hat{a}.
\end{align}
Let's now consider the evolution of a generic Fock state $\ket{m}$ of the QB under a single collision with a coherent ancilla, initialized as Eq.~(\ref{eq:anc_dm}) with $c=1$. As it was first pointed out in Ref.~\cite{Slosser_89} the probability amplitudes of finding the oscillator in $\ket{m-1}$ or in $\ket{m+1}$ after a collision are given respectively by 
\begin{equation}
\begin{split}
    P_{m\rightarrow m-1} &= \mathrm{Sin}(\theta \sqrt{m}),\\
    P_{m\rightarrow m+1} &= \mathrm{Sin}(\theta \sqrt{m+1}).
\end{split}
\end{equation}
For specific $m_l$ satisfying
\begin{equation}
\label{eq:trapping_condition}
    \theta\sqrt{m_l} = l\pi, \quad l \in \mathbb{Z}
\end{equation}
one has 
\begin{equation}
    P_{m_l\rightarrow m_l-1} = P_{m_l-1\rightarrow m_l} = 0.
\end{equation}
The corresponding state $\ket{m_l}$ is known as a \textit{downward trapping state}, since it is decoupled from lower Fock states, whereas $\ket{m_l-1}$ as an \textit{upward} one, being decoupled from Fock states above it. These two states then constitute a pair of uncoupled states~\cite{Slosser_89,Nemeth_05}. It is worth noting that the only possible state satisfying Eq.~(\ref{eq:trapping_condition}) with $l=0$ is the ground state, which therefore can be seen as a natural \textit{downward trapping state}.\\
Eq.~(\ref{eq:trapping_condition}) implies
\begin{equation}
\label{eq:trap_numbers}
    m_l = \qty(\frac{l\pi}{\theta})^2,
\end{equation}
for this to be an integer for each possible $l$, one needs to fine-tune the coupling to
\begin{equation}
    \theta = \theta_{ft} = \frac{\pi}{\sqrt{s}}, \quad s \in \mathbb{N}^{*}
\end{equation}
which is the condition introduced in the main text. This ensures that 
\begin{equation}
\label{eq:trap_integers}
    m_l = l^2 s
\end{equation} 
are all integers and that there exists an entire set of Fock pairs, labelled by $l$, formed by an \textit{upward trapping state} and the corresponding \textit{downward one}. Due to this fact, the oscillator Hilbert's space is divided into dynamically disconnected blocks, labeled by $l$, each of which is determined by the Fock states with $m_{l} \leq m < m_{l+1}$~\cite{Slosser_89,Nemeth_05}. \\
What discussed so far implies that, if the oscillator is initialized in a given block, it will remain there indefinitely under the JC collision model, leading to the so called \textit{trapping dynamics}. It can be directly derived from Eq.~(\ref{eq:trap_integers}) that the size of each block is
\begin{equation}
    d_{l+1} \equiv m_{l+1}-m_{l} = (2l+1)s.
\end{equation}
In this way, the block containing the ground state, namely that with $l=0$, has a size $d_1=s$. Since the QB starts in the ground state, the fine-tuning condition of Eq.~(\ref{eq:theta_ft}) controls the portion of the oscillator Hilbert space which is accessible during the charging protocol in the JC regime. To better clarify the situation let's consider as an example the case $\theta = \frac{\pi}{\sqrt{3}}$ ($s=3$). Here, the blocks formed in the QB Hilbert space have sizes $\qty(d_1=3,d_2=9,d_3=15 \dots)$ and the lower block, generated by the Fock states $\qty{\ket{0},\ket{1},\ket{2}}$, will be the only part of the Hilbert space dynamically accessible for a QB initialized in the ground state. \\
This feature of the JC Markovian Micromaser dynamics also holds in absence of quantum coherences at the level of the ancillas ($c=0$)~\cite{Shaghaghi_MM_23,Nemeth_05} and has important consequences for the charging of a QB initialized in its ground state. Indeed, it imposes a stringent upper bound for the maximum energy storable via a fine-tuned collisional charging protocol starting from the ground state, namely $E_{max}=(s-1)\omega$. Moreover, according to what discussed in the main text, the emergence of the trapping chambers is not affected by non-Markovian effects as long as the battery-memory and the memory-chargers interactions occurs in separated steps.

\section{Determining the steady state of a composite CM}
\label{Steady}

In order to determine the steady state of the QB, we start by considering the fact that the dynamics of the enlarged $Q+M$ system is a Markovian sequence of identical maps $\mathcal{N}$~\cite{Ciccarello21}. By determining its steady state and tracing out the memory degrees of freedom from this, we are able to recover the steady state of $Q$. The map $\mathcal{N}$ is obtained following the two-fold structure of the single collisional step as described in Sec.~\ref{sec:model}. If we denote with $\hat{\uprho}_{Q,M}$ the density matrix before the $n$--th collisional step, the map can be expressed as
\begin{align*}
\mathcal{N}[\hat{\uprho}_{Q,M}] &= \mathrm{Tr}_k \bigl\{ \hat{U}_{M,A_k} \hat{U}_{Q,M} \hat{\uprho}_{Q, M} \otimes \ket{+}\bra{+} _{k}\hat{U}^\dagger_{Q,M} \hat{U}^\dagger_{M,A_{k}} \bigr\} 
\end{align*}
where we have replaced the density matrix of the ancillas with the corresponding pure state $\ket{+}_k$. Performing the trace over a basis formed by $\ket{+}_k$ and a TLS state orthogonal to it, namely $\qty{\ket{+}_k,\ket{-}_k}$, one obtains
\begin{equation}
\begin{split}
\label{eq:composite_CM_Kraus_dec}
\mathcal{N}[\hat{\uprho}_{Q,M}] &=  \bra{+}_k \hat{U}_{M,A_{k}} \ket{+}_k \hat{U}_{Q,M}\hat{\uprho}_{Q,M} \hat{U}_{Q,M}^\dagger \bra{+}_k \hat{U}^\dagger_{M,A_k} \ket{+}_k +\\
& +\bra{-}_k \hat{U}_{M,A_{k}} \ket{+}_k \hat{U}_{Q,M}\hat{\uprho}_{Q,M} \hat{U}_{Q,M}^\dagger \bra{+}_k \hat{U}^\dagger_{M,A_{k}} \ket{-}_k \\
&= \sum_{\alpha=+,-}^{} \hat{K}_{\alpha+}\, \hat{\uprho}_{Q,M}\, \hat{K}_{\alpha+}^\dagger
\end{split}
\end{equation}
where we defined the Kraus operators $\hat{K}_{\alpha+}$ as 
\begin{equation}
    \hat{K}_{\alpha+} = \bra{\alpha}_k \hat{U}_{M,A_{k}} \ket{+}_k \hat{U}_{Q,M}.
\end{equation}
Eq. (\ref{eq:composite_CM_Kraus_dec}) can now be vectorized as~\cite{Havel_03}.
\begin{equation}
\label{eq:superoperator_vectorized_composite}
\mathrm{\textbf{col}}\!\left(\mathcal{N}[\hat{\uprho}_{Q,M}]\right) = \left( \sum_{\alpha} \hat{K}_{\alpha+}^* \otimes \hat{K}_{\alpha+} \right) \times \mathrm{\textbf{col}}\!\left(\hat{\uprho}_{Q, M}\right) \equiv \mathcal{\textbf{\textit{K}}}
\times\mathrm{\textbf{col}}\!\left(\hat{\uprho}_{Q, M}\right).
\end{equation}
It is worth noting that the superoperator $\mathcal{\textbf{\textit{K}}}$ acts on vectors with dimension $(2d)^2$, with $d$ the size of the truncated Hilbert space of $Q$. \\
Diagonalizing $\mathcal{\textbf{\textit{K}}}$ and finding the eigenstate with eigenvalue $\lambda=1$ provides us with the steady state of the map $\mathcal{N}$, which corresponds to the steady state of the Markovian dynamics of $Q+M$. By performing a partial trace over the $M$ degree of freedom one gets to the steady state of the non-Markovian Micromaser.

\bibliographystyle{unsrt}
\bibliography{references}

\end{document}